\documentclass[pra,aps,preprint]{revtex4}

 \usepackage{graphicx}
 \usepackage{dcolumn}
 \usepackage{bm}
 \usepackage{subfigure}
 \newcommand{\YSO}{Y$_2$SiO$_5$}

\begin{document}

 \title{Stopped light with storage times greater than one second
   using EIT in a solid}

 \author{J. J. Longdell}
 \email{jevon.longdell@anu.edu.au}
 \author{E. Fraval}
 \author{M. J. Sellars}
 \author{N. B. Manson}
 \affiliation{Laser Physics Centre, Research School of Physical
    Sciences and Engineering, Australian National University,
    Canberra, ACT 0200, Australia.}
\begin{abstract}
  We report on the demonstration of light storage  for times greater
  than a second in praseodymium doped Y$_2$SiO$_5$ using
  electromagnetically induced transparency. The long storage
  times were enabled by the long coherence times possible for the
  hyperfine transitions in this material.  The use of a solid state
  system also enabled operation with the probe and coupling beam
  counter propagating, allowing easy separation of the two beams. The
  efficiency of the storage was low because of the low optical
  thickness of the sample, as is discussed this deficiency
  should be easy to rectify.
\end{abstract}

  \pacs{42.50.Gy,42.62.Fi }
  \keywords{Quantum memory, Electromagnetically induced
    transparency,slow light, Coherent Spectroscopy, Rare-earth}
\maketitle



Some of the most significant advances in quantum information
processing have been made using quantum optics-based techniques.  For
example, working practical quantum cryptosystems already exist and
there have been demonstrations of linear optics quantum computing
\cite{obri03}, quantum teleportation, quantum non-demolition
measurements \cite{Pryd04}, quantum feedback and control
\cite{stoc04}. To proceed further it is necessary to have devices such
as single photon sources, quantum memories and quantum repeaters,
where quantum information is exchanged in a controlled fashion between
light fields and material systems. It has been proposed that both the
required control and strong coupling can be readily achieved using an
ensemble approach, where the light field interacts with a large number
of identical atoms. Such a ensemble based approaches now exist for single
photon sources \cite{dlcz}, `cat' state sources \cite{pate03}, quantum
memories \cite{flei02, mois01, eisa04} and quantum repeaters
\cite{dlcz}.  Experiments have demonstrated heralded single photon
sources \cite{kuzm03, jian04} and the mapping of quantum information
on a light field onto spin states of an atomic ensemble \cite{Juls04}.
Experiments using electromagnetic induced transparency have
demonstrated the storage and recall of optical pulses \cite{liu01,
  phil01}.

The quantum systems used for these ensemble based demonstrations have
almost exclusively been atomic vapors. An issue with these
demonstrations is that even for a laser cooled ensembles, the atomic
motion impacts on the devices' performance. Ensembles of solid-state
optical centers provide an alternative to atomic systems where the
relative motion is zero.  In this paper we investigate the use of
solid-state system for ensemble based quantum optics and highlight its
usefulness by stopping a light pulse using electromagnetically induced
transparency (EIT). Unlike an earlier experiment \cite{turu02}, the
current demonstration highlights for the first time two advantages of
using optically active solid state centers: a one thousand fold
increase in storage time and the ability to operate with a less
restrictive beam geometry.

When storing light using EIT characteristics of the field are recorded
as a spin wave in the ensemble. The storage time is determined by the
coherence times of the hyperfine transitions. In principle coherence
times for hyperfine transitions in atomic systems can be very long and
many minutes have been measured in ion traps \cite{fisk95}. However,
these long coherence times in large ensembles suitable for EIT have
not been achieved. Transit time broadening in vapor cells and magnetic
inhomogeneity in trapped systems mean that the longest that light has
been stored atomic systems is a few  milliseconds. In contrast, in
earlier work we have demonstrated techniques to obtain hyperfine
coherence times of tens of seconds in Pr:\YSO\ \cite{frav04a,frav04b}.
Here we show it is possible to utilize these long coherence times to
stop light for similar lengths of time.

EIT is sensitive to atomic movement, with the spin wave being
scrambled once the atoms have moved significantly compared to the
wavevector mismatch between the probe and the coupling beams. To
minimize this wavevector mismatch experiments in atomic systems
typically operate with the beams co-propagating. Because the probe
and coupling beams are close in frequency, in this configuration,
the wavevector mismatch is typically less than 1~cm$^{-1}$. A
consequence of this co-propagating operation is that it is
difficult to separate the probe and the coupling beam. In a
solid-state system, where the optical centers are locked in a
crystal lattice, co-propagating operation is not required, in the
present work the probe and the coupling beams are
counter-propagating. With counter-propagating beams it is easier
to separate the probe and coupling beam whilst maintaining optimum
overlap.

\begin{figure}
  \includegraphics[width=0.75\textwidth]{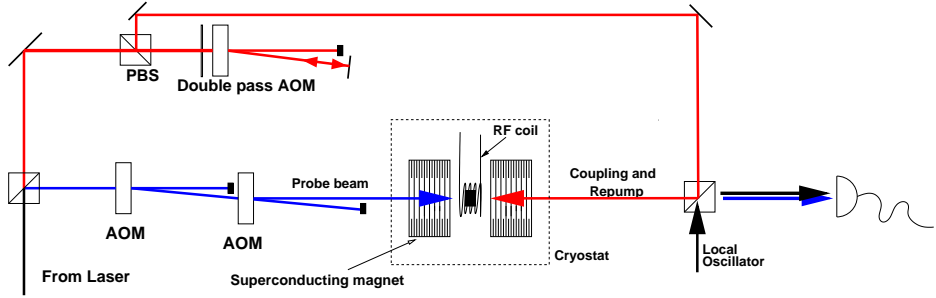}
  \caption{The experimental setup. What is not shown is a beam picked
    of the laser was put in the remaining port of the right most
    beam-splitter. This enabled heterodyne detection of the probe beam
  that was transmitted through the sample.}
  \label{fig:setup}
\end{figure}

The experimental setup is shown in FIG.~\ref{fig:setup}. Because
of the narrow 2500~Hz optical homogeneous linewidth of the $^3$H$_4$$\rightarrow$$^1$D$_2$
transition in Pr$^3+$:Y$_2$SiO$_5$ a highly frequency stabilized
dye laser was required for the experiment not to be limited by
laser jitter. The laser used was a modified Coherent 699 dye laser
with a linewidth 200~Hz over 1~second time scales. The laser
output was split into two beams, one of which was frequency
shifted and gated by two AOMs and used as the probe beam. The
other beam was frequency shifted and gated using a double pass AOM
setup. This beam was used for the coupling and repumping fields. This coupling/repumup beams was aligned on a beam-splitter to go through the sample counter-propagating with the probe. The
spare port of this right-most beam splitter was used to combine a local
oscillator beam with the transmitted probe beam,
enabling the heterodyne detection of the signal.

The sample used was the same as that used in reference \cite{frav04b}
and consisted of 0.05\% Praseodymium doped in Y$_2$SiO$_5$. It was
4~mm thick along the direction of light propagation.  The sample was
mounted in a bath liquid helium cryostat. Three orthogonal
super-conducting magnets were used to apply a DC magnetic field to the
sample and a six turn rf coil was used to apply a rf field.

The dominant dephasing mechanism for the hyperfine states of the
Pr$^{3+}$ ions is random Zeeman shifting due to fluctuating
magnetic fields from the yttrium nuclei. Dramatic increases in
coherence times can be achieved by operating at a magnetic field
where the transition frequency is insensitive to magnetic field
changes to first order \cite{frav04a}. The magnetic field required
is 78~mT in an  orientation described in Ref.~\cite{frav04a}.
Once the magnetic field is obtained  the
remaining fluctuations have reasonably long correlation times.
This situation enables the effective use of dynamic decoherence control
(DDC) techniques \cite{viol04} and coherence times in excess of
half a minute have been demonstrated \cite{frav04b}.


\begin{figure}
  \includegraphics[width=0.3\textwidth]{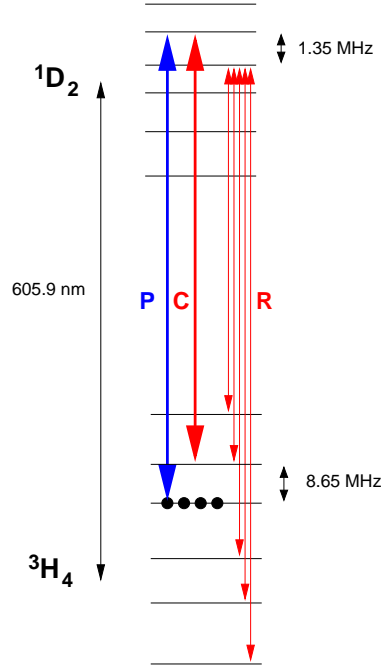}
  \caption{Energy level diagram showing transitions driven as part of
    the experiment. The experiment was carried out on the zero phonon
    line of the $^3$H$_4$$\rightarrow$$^1$D$_2$ optical
    transition. The hyperfine levels are shown and these are due to
    the $5/2$ spin of the praseodymium nuclei. In the presence of a
    magnetic field these are linear combinations of the zero field
    states. The probe, coupling and repump beams are labelled
    P, C and R respectively.}
  \label{fig:nrg_levels}
\end{figure}

An energy level diagram showing the transitions driven during the
experiments is shown in Fig.~\ref{fig:nrg_levels}. While the optical
inhomogeneous line widths is a few GHz. The narrow homogeneous
linewidth (of order 1~kHz) and long hyperfine population lifetimes (of
order 1~minute) enabled the experiment to be carried out on an
ensemble with a much smaller range of optical frequencies. At the
beginning of each shot a sequence of the five optical frequencies
(labelled ``R'' in Fig.~\ref{fig:nrg_levels}) was applied
repeatedly. The repump frequencies were applied sequentially rather
that all at once to avoid the possibility of darks states and
nonlinear mixing of the different frequencies in the AOM. The gap in time
between the repumping and each experimental shot was long enough to
ensure that ions had no  remaining optical coherence.
This repumping procedure prepared an ensemble of ions in the desired
hyperfine state and gives a narrow adsorption with an inhomogeneous
width of 100~kHz when measured by sweeping a week probe in frequency
(line given by dots in Fig.~3). When the coupling beam was applied a
narrow transparency was obtained in the absorption of the weak probe
(solid trace in Fig.~3).


\begin{figure}
  \centering
  \includegraphics[width=0.4\textwidth]{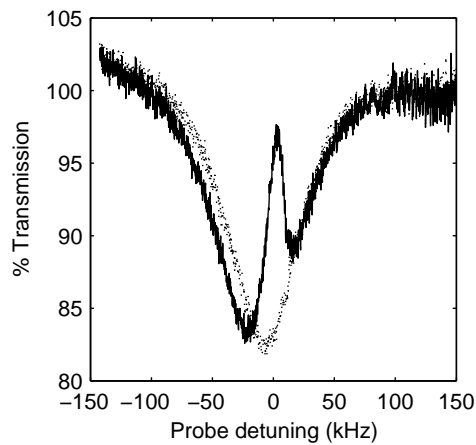}
  \caption{Transmission of a weak probe, 10~$\mu$W, as its frequency is swept
    through resonance with the prepared ensemble. The solid line was
    taken with a 1~mW coupling beam on and the dotted line with the
    coupling off.}
  \label{fig:eit_sweep}
\end{figure}


 The repumping
beams were applied after each shot and the 300~kHz span shown was
swept in 4~ms. The transmitted probe beam was detected as a
heterodyne beat signal and the bandwidth of the RF detector was
comparable to 300~kHz, the extra noise at each end of the spectrum
came from dividing out this frequency response. For coupling
intensities above  1~mW the EIT was observed to depend linearly
on the amplitude of the coupling beam. The limiting EIT width at
low intensity was 10~kHz, corresponding to the hyperfine
inhomogeneous linewidth. Below 1~mW the EIT transmission decreased
with decreasing coupling intensity.

It can bee seen from FIG.~\ref{fig:eit_sweep} that the peak
absorption of our ensemble is only about 15\% and, as is discussed
below, this limits the efficiency of the storing process.


\begin{figure}
  \centering
  \includegraphics[width=\textwidth]{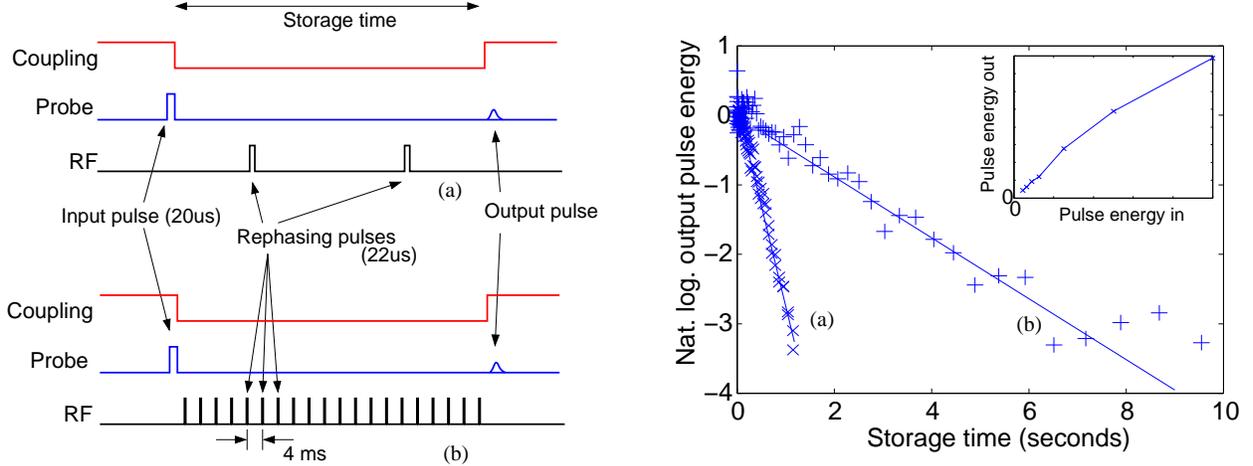}
  \caption{ On the left is the time sequence used in the stopped light
    experiments (a) with simple rephasing of the inhomogeneous
    broadening of the spin transitions and (b) with ``bang-bang''
    dynamic decoherence control. In (a) two rephasing pulses are used
    placed 1/4 and 3/4 of the way through the storage time. In (b) N
    rephasing pulses were used (N even). The first rephasing pulse was
    applied 2~ms after the light was stored, the pulses were separated
    by 4~ms and the last pulse was applied 2~ms before the light was
    recalled. The rephasing pulses lasted 22~$\mu$s.  On the right is
    the size of the recalled pulse as a function of time. The
    faster decay was acquired using simple rephasing of the ground
    state spin coherence (a). The slower decay was acquired using
    ``bang-bang''\cite{viol98} decoherence control (b). The inset
    shows the energy of the recalled pulse as a function of the
    energy of the input pulse, the probe pulse length was 20~$\mu$s
    and the delay held constant at 100~ms.  }
  \label{fig:superfig}
\end{figure}

The time sequence for the light storage demonstration is shown on the
left of 
FIG.~\ref{fig:superfig}. A 20~$\mu$s long probe pulse was
applied and then the 10~mW coupling beam was turned off to
transfer the optical coherence onto the spin transition. As in the
previous solid state stopped light \cite{turu02} experiment RF
rephasing pulses were used to rephase the inhomogeneous broadening
in the spin transition. Although one RF rephasing pulse is enough
to rephase the spin-wave it also flips the spin-wave's direction.
Therefore when not using co-propagating beams, as is the case
here, it is necessary to use an even number of rephasing pulses.

The size of the pulse of light recalled as a function of delay can be shown with and without dynamic
decoherence control (DDC) and the results are shown in FIG.~\ref{fig:superfig}. The decay constants for the stored
signal output were 0.35 seconds without DDC and 2.3 seconds with
DDC. These decay rates were comparable to measurements of $T_2$
made using the same method as Fraval et al. \cite{frav04a}. The
difference between the present measurements of $T_2$ and those
obtained by Fraval et al. \cite{frav04a} is attributed to not
having tuned the magnetic field as carefully as was achieved by Fraval et al.

Shown in the inset of FIG.~\ref{fig:superfig} is the intensity of the output
pulse as the intensity of the input probe pulse is varied. From
the graph it can be seen that the storage process is linear at low
powers and starts to saturate at higher powers once the input
pulse becomes a significant fraction of a $\pi/2$ pulse. This
demonstration of linearity is important. Previous solid state
experiments \cite{turu02} have been restricted  by laser frequency
jitter to using probe pulses with areas greater than $\pi$. At
such high powers effects such as self induced transparency (SIT)
\cite{mcca69} cannot be ignored.

While the effect was linear and scaled to low powers, the
efficiency was low, of the order of 1\%. This in part can be
improved with better timing and shaping of the probe and coupling
waveforms. However the main reason for the low efficiency is the
low optical absorption at the probe frequency and the accompanying
modest group delay. 


The sample used for this experiment was only 4~mm thick,
longer samples as well as multi-pass cells and cavities are simple
means to increase the optical absorption. Preliminary measurements on
a samples with a range of praseodymium concentrations \cite{ala}
suggest that at least two or three fold increases in the optical
thickness can be achieved by increasing the concentration without
significantly increasing the inhomogeneous broadening of the hyperfine
transition.

As it is a goal of this line of research to store and retrieve quantum
mechanical states it worthwhile to consider the effect that rephasing
pulses would have on few photon states stored in the hyperfine
coherences. It has been asserted in a theoretical investigation of
quantum information storage in the solid state \cite{john04} that one
would not be able to apply the RF $\pi$ pulses with sufficient
accuracy. This is not the view of the authors of this paper. In
Ref.~\cite{john04} it was assumed that the $\pi$ pulse would
have to be applied with an accuracy close to 1 part in $N$ (where $N$
is the number of atoms) in order that the few photon pulse not be
swamped by light caused by inaccuracies of the $\pi$ pulse. However
this light will be emitted randomly rather than in the very precise
spatio-temporal mode of the output pulse. This should enable the
output pulse to be easily separated from the background with very high
efficiency. 


In conclusion, we have demonstrate stopped light in Pr:\YSO\ for time
scales of several seconds which is three orders of magnitude longer
than any obtained previously. Based on previous measurements of $T_2$
it should be possible to extend this storage time by at least one more
order of magnitude.

For the first time stopped light has been demonstrated in a solid with
the coupling and probe beams counter propagating.  This configuration
is desirable as it allows easy separation of the two beams.  However,
it is only practical if the atoms movement during the storage time is
small compared to the optical wavelength.  Even for ultra-cold systems
this places significant limits storage time.  In a solid where the
atoms are locked into position this isn't a problem.

The efficiency of the storage process required for a quantum memory
should be obtainable by increasing the density of the dopant ions and
by increasing the interaction length.

The authors would like to thank Philip Hemmer for helpful discussions.
We would like to acknowledge the support of the Australian Research Council and the Australian Department of Defense.

\end{document}